# Characterizing the neural correlates of reasoning

*David Papo[1,2]\**

[1] *Laboratory of Biological Networks, Center for Biomedical Technology, Universidad Politécnica de Madrid*

[2] *LPC Institute, Madrid*

**\*Correspondence:**
*David Papo, Laboratory of Biological Networks, Center for Biomedical Technology, Universidad Politécnica de Madrid, Calle Ramiro de Maeztu, 7, 28040 Madrid, Spain*
*e-mail: papodav@gmail.com*
*www: https://davidpapo.wordpress.com/*

The brain did not develop a dedicated device for reasoning. This fact bears dramatic consequences. While for perceptuo-motor functions neural activity is shaped by the input's statistical properties, and processing is carried out at high speed in hardwired spatially segregated modules, in reasoning, neural activity is driven by internal dynamics, and processing times, stages, and functional brain geometry are largely unconstrained *a priori*. Here, it is shown that the complex properties of spontaneous activity, which can be ignored in a short-lived event-related world, become prominent at the long time scales of certain forms of reasoning which stretch over sufficiently long periods of time. It is argued that the neural correlates of reasoning should in fact be defined in terms of non-trivial generic properties of spontaneous brain activity, and that this implies resorting to concepts, analytical tools, and ways of designing experiments that are as yet non-standard in cognitive neuroscience. The implications in terms of models of brain activity, shape of the neural correlates, methods of data analysis, observability of the phenomenon and experimental designs are discussed.

## INTRODUCTION

Consider an individual trying to solve a problem, and reasoning for ten minutes before attaining a solution. Take the middle five minutes. Clearly, though containing no behaviourally salient event, these five minutes represent a genuine, indeed rather general, instance of reasoning. What do we know about the brain regime far from its conclusion? Can we use this regime to predict a solution, and a solution to retrodict this regime?

Here, I concentrate on a form of reasoning, of which the above scenario constitutes an example, which can broadly be defined as "thinking in which there is a conscious intent to reach a conclusion and in which methods are used that are logically justified" [1], with no a priori assumption on the type of reasoning process that may take place during it. It is argued that finding the *generic* properties of this form of reasoning entails addressing the following fundamental issues: What are reasoning's temporal and spatial scales? When is a given observation time sufficient? How should we integrate the information contained in various reasoning episodes?

## A MINI LITERATURE REVIEW

The neural correlates of reasoning have traditionally been expressed in terms of brain spatial coordinates. Early neuropsychological work viewed reasoning as emerging from global brain processing [2], consistent with evidence indicating that it is negatively affected by diffuse brain damage [3]. Neuroimaging studies have framed the neural correlates of reasoning in terms of local functionally specialized brain activity, either by taking a normative approach to reasoning [4-10], or by fractionating it into sub-component processes [11-14]. The results often lack specificity to reasoning [15]. Most importantly though, these investigations provide a static characterization of reasoning.

The neuroimaging literature mostly focused on short-term and normative forms of reasoning [9,16-18]. This minimizes variability in reasoning episode length and allows segmenting reasoning episodes into separable chunks, but does that at the price of limitations in the phenomenology and ecologic value of its stimuli. Some neuroimaging [19,20] and electrophysiological [21-29] studies examined more ecological forms of reasoning, viz. insight problems [30]. However, even electrophysiological studies, despite optimal temporal resolution, adopted an event-related perspective, concentrating on activity occurring few seconds before insight emergence, which only documents the *outcome* of the reasoning process, not the process itself.

Event-related neural activity associated with the solution of riddles with insight was found to be related to properties of preceding resting activity [26,27]. These studies had the remarkable merit of using spontaneous brain activity to characterize reasoning, but in essence provided a comparative statics description. Although some behavioural studies treated reasoning as a dynamical process [31], a comparable neurophysiological characterization is still incomplete. Altogether, the research accomplished so far has generally not looked at reasoning as a dynamical process, and produced either time-averaged frames or discrete maps in partial time cuts.

## THE PROBLEM(S) WITH REASONING

The generalized form of reasoning considered in this study comes in episodes offering scant behaviourally salient events with no characteristic temporal length. Each episode is a non-reproducible instance, as reasoning task can be carry out in multiple ways. Brain activity associated with reasoning is not event-related, and many neurophysiological processes interact in a wide range of spatial and temporal scales.

These phenomena can all be traced back to a basic fact: the brain did not develop a dedicated device for reasoning. Hardwired partially segregated modules ensure that perceptuo-motor functions are carried out at great speed, with stereotyped duration and time-varying profile, and identifiable stages, largely determined by input statistical properties. Reasoning, on the contrary, is an internally-driven dynamics: processing times, stages, and functional brain geometry are largely unconstrained.

Considering these extraordinary challenges, can we still find general reasoning properties, over and above specific task demands and individual differences? What sort of process is reasoning in its general form? Is it a series of simpler reasoning cycles? Can we segment it into stages? What are the best neural variables and tools to make these properties observable?

## CHARACTERIZING THE REASONING PROCESS

Robust characterizations of reasoning should incorporate *stylized facts*, i.e. properties consistently appearing on different subjects and in different periods of time, and select analytical tools accordingly. For instance, perceptual response sensitivity to incoming signals, stability against noise, and minimal dependence on initial conditions favour tools capturing transient dynamics, which naturally reproduce these properties under appropriate conditions, over tools handling asymptotic activity, which fail to do so [32].

Reasoning's relative instability and inefficiency suggest that optimal circuitry may need constant reconstruction and protection from interference, summoning protracted support of energetically costly long-range communications. Thus, reasoning may be a sort of resonant regime, where functional efficiency would be achieved with specific, though unstable, spatio-temporal patterns, and should be studied with tools capable of extracting spatially-extended dynamic transients.

## REASONING DYNAMICS

Each cognitive process can be translated in dynamical terms and corresponding aspects of neural activity.

Perceptual processes are relaxational, quasi-stereotyped short duration processes. The brain can *prima facie* be modelled as an excitable medium: perturbations above a threshold induce a dynamical cycle, before the system reverts to its initial silent state.

Learning too is a relaxational process: following a gradient dynamics, the brain incorporates the environment's statistical relationships, by representing them in terms of its functional connectivity. Cycles can be of much longer duration and non-trivial shape than perceptual ones. The dynamics is dominated by fluctuations much shorter than the whole process.

Reasoning may not be purely relaxational. The corresponding neural activity is a fluctuation-dominated endogenously modulated spontaneous brain activity, with no clear gradient, and no single instant summarizing the entire process. The reasoning scientist's world is considerably more complex than the event-related short time scale one of perception. It is through the generic properties of spontaneous activity's long time scales that robust reasoning properties should be formulated [33].

### THE STARTING POINT: SPONTANEOUS BRAIN ACTIVITY

Spontaneous activity can be thought of as a data bank of cortical states, continuously reedited across the cortex [34]. This re-editing process contains rich non-random spatio-temporal structure [35-40].

The building blocks of this structure are fluctuations which the brain, as all dissipative out-of-equilibrium systems, generates even for fixed control parameter values and in the absence of external stimuli, and which constitute the trademark of its functioning. The dynamics is intermittent, with alternating laminar and turbulent phases [41,42], weakly *non-ergodic*, i.e. some phase space regions take extremely long times to be visited, and shows *aging*, i.e. temporal correlations depend on the observation time [43]. Various aspects of spontaneous activity display similar properties at all temporal and spatial scales [44-51]. Self-similarity break-down [52,53], with Gaussian low-frequency and non-Gaussian high-frequency fluctuations [54] were also reported.

### Understanding fluctuations

We can imagine brain activity as the motion of a random walker, making steps of a length taken from some distribution, at times taken from some other distribution or, equivalently, of a macroscopic particle diffusing in a liquid, subject to viscous friction and to an additive random force [55].

The relationship between these two forces' time scales determines how microscopic fluctuations produce observable macroscopic properties. In the equilibrium world of perceptual scientists, the brain makes steps taken from a Gaussian distribution and whose correlations, much faster than the friction time scale, have no macroscopical effect. Reasoning scientists live far from equilibrium: random fluctuations are no faster than the friction term and show long-lasting correlations, which renormalize, becoming macroscopically detectable [56].

Complex fluctuations reveal the particle's navigation 'style', e.g. how travelled distances and the times to reach a given target scale with time [57]. They also allow deducing a system's characteristic temporal scales. For an equilibrium system with exponential temporal autocorrelation decay, the *correlation length*, i.e. the value $\xi$ that makes the autocorrelation $C(t=\xi)=0$, or the *correlation time* $\tau_C = \int_0^\xi C(t)dt$, endow the process with a unique temporal scale. However, at long time scales, the presence of scaling indicates that the brain relaxes more slowly than an exponential and fluctuates at all scales [58-60]. Both $\xi$ and $\tau_c$ may diverge, and a characteristic time ceases to exist. Temporal scales are characterized by some relationship between them, which can be treated as a dynamical system, relating the behaviour of an observed variable over a series of nested scales [61,62].

The brain's functional and corresponding dynamical heterogeneity produces a spatial distribution of time scales. While only some of these are usually considered of interest, global temporal scales may not coincide with any of the local dynamics, and may emerge from transient connectivity patterns created and destroyed by rewiring processes [63]. Because spatial scales also have non-trivial topological properties [64], global dynamics is a field endowed with arbitrarily real and phase-space complex topologies [65], where scales replace metric distances, fractal geometry the Euclidean one, and scale invariance Galilean invariance.

By resorting to nonlinear analysis, algebraic and differential topology, renormalization group methods etc. [66,67], brain activity can then be described in terms of universal properties, i.e. regimes with robust macroscopic behaviour with respect to the nature of microscopic interactions and sharing the same symmetries. These descriptions partition the phase space, identify dynamical pathways leading to specific regions of this space, and allow relating descriptions of the same brain at different scales, and grouping descriptions of different brains exhibiting the same large-scale behaviour [66].

### Chunking

Correlated noise and cross-scale relationships produce *temporally ordered* structures which can help segmenting reasoning episodes into chunks.

This can be done by defining quasi-stationary segments boundaries [68]. The scaling properties of quasi-stationary segments' durations may help clarifying whether reasoning in its general form is merely a repetition of simple cycles seen in more controlled forms of reasoning, or is of a qualitatively different nature, and in any case, determining the time scales at which simpler cycles are reedited.

The waiting-time distribution between steps defines an internal *operational time*, which may grow sub- or super-linearly with *physical time* [69]. Multiplicative cross-scale interactions bias the waiting-time distribution so that *operational* and *physical* times no longer coincide, and local probability densities become time-dependent and intermittent [70].

### FROM SPONTANEOUS ACTIVITY TO REASONING

Cognitive processes can be thought of as selections and orchestrations of cortical states already present in spontaneous activity [71,72]. Each process corresponds to a specific phase space cut, with its own topological properties and symmetries, and characteristic kinematics, memory, ageing properties, degree of ergodicity, and internal clock [33].

Reasoning may modulate not brain activity's frequency or amplitude but its functional form [33], e.g. by pushing it towards the basin of attraction of advantageous probability distributions: good reasoning could be tantamount to designing a driving noise function forcing the system's stationary distribution to equal a target one.

Cognitive demands may generally change the symmetries of brain activity. For instance, scaling regime modulations observed in associations with a reasoning task [73] may correspond to cross-overs between universality classes, reflecting dynamical transitions in the system's behaviour [74].

### APPRAISING REASONING

Brain fluctuations can be interpreted in various interrelated ways that help evaluating the *quality* of reasoning, by using models of the function it fulfils and quantifying the constraints the brain faces while performing it.

### Metaphors for reasoning

Reasoning, as other cognitive processes, e.g. memory recall [75,76], can be represented as a search process similar to that of animals foraging in an unknown environment [77]. This search process can be characterized in terms of random walks [78-81]. Importantly, random walk *types* can quantify the extent to which a given trajectory optimizes search, given the characteristics of the explored space and the resources available to the individual [81]. Such a characterisation would allow assessing in a context-specific way the quality of both the reasoning and the 'reasoner'. That behavioural aspects of human cognition [75,76] and brain activity both show non-Gaussian, heavy-tailed distributions might indicate search optimality [80,82]. However, because these properties are generic in spontaneous activity, reasoning's quality can only be described in terms of its modulations. Furthermore, since local dynamics may be nowhere Lévy-like, finding the neural property and spatial scale showing scaling are the crucial steps.

The reasoning regime could also be represented as a network traffic regulation problem, where phenomena such as overload or jamming may be quantified in terms of information creation, erasure and transmission rates, by regarding simple fluctuations as letters of an alphabet and fluctuation complexes as words, and quantifying the amount of information in the system. Characterizing traffic regulation may involve understanding the interplay between the underlying network's topology, burstiness of information packets and the shape of fluctuation distributions [83-85]. Although only causal information [86] may directly serve reasoning purposes, the total information encoded in the network may describe the noise-control mechanisms indirectly optimizing it.

The sudden onset of insight may be thought of as an extreme event comparable to earthquakes, financial crashes or epileptic seizures [87,88], e.g. as a rupture phenomenon, and the route to it as a long charging process, with nested hierarchical "earthquakes", and try predicting its occurrence. For such phenomena, the coupling strengths distribution and topology constitute the relevant field [89], which may be described using complex network theory [64]. It is tempting to conjecture that insight onset may be predicted by monitoring e.g. anomalous diffusion parameters [88], variations in Gaussianity [90], or changes in fractal spectrum complexity [91,92].

### From dynamics to thermodynamics

Differences in reasoning abilities can produce not only differences in processing times of some orders of magnitude, but also qualitative ones in dynamical and statistical aspects of brain fluctuations, and in corresponding brain topography and topology. A natural question when studying reasoning is then: "How efficiently does a given brain carry out reasoning?".

There are various ways to assess efficiency of a given device. If reasoning was the output of an engine, e.g. a Carnot's engine, it would be natural to understand not only how the engine produces it, but also how efficiently it performs it.

While functional network reconstruction can provide an indirect characterization of brain activity's energetic aspects [93,94], the brain's thermodynamics can directly be deduced from its dynamics [95], which can be interpreted as the walker's diffusion on the surface of the entropy production rate in the space of kinetic variables [96].

Temperature represents a good example of how thermodynamical variables can be used to describe brain activity [97]. For a system at equilibrium, temperature is proportional to the ratio between the response to an external field conjugate to some observable and the corresponding autocorrelation function in the unperturbed system [98]. In the brain [99], equilibrium temperature must be substituted by an *effective temperature* [100] reflecting what a thermometer responding on the time scale at which the system reverts to equilibrium would measure [101]. Each scales can have its own effective temperature even within the same spatial region, corresponding to different distances from equilibrium and reflecting qualitatively different diffusion processes [102]. Thus, measuring effective temperature at various scales allows understanding the extent to which each spatio-temporal scale deviates from equilibrium, produces entropy, etc..

Thermodynamic functions can be used to directly describe brain activity, but also as control parameters, i.e. one can monitor different aspects of brain activity as one measured thermodynamical variable varies in time. For instance, one may observe temperature variations during a reasoning task, but also possible phase transitions in some other property of neural activity, as temperature is varied [97].

## FROM THEORY TO EXPERIMENT

### OBSERVING REASONING

Reasoning is a difficult phenomenon to observe: tasks can be executed in more than one way, each possibly corresponding to a neural phase space with convoluted geometry and the processes involved in reasoning may evolve over time-scales exceeding those typical of laboratory testing.

Proper observation of a given process requires that the *observation time* be much larger than any scale in the system. A process is observable if it has a finite ratio between the characteristic time of the independent variable and the length of the available time series [103]. Factors including long-term memory, aging and weak ergodicity breaking may result in a diverging ratio [104]. The observation time should also be much larger than the time needed to visit the neural phase space. However, deciding whether a reasoning episode, or even an ensemble of episodes, sufficiently sample a subject's repertoire is non-trivial.

Cognitive neuroscientists observe phenomena through experiments where subjects typically carry out given tasks a large number of times, assumed to be independent realizations of the same observable, and to adequately sample the phase space of task-related brain activity. However, in the presence of complex fluctuations, trials may not *self-average*, i.e. dispersion would not vanish even for an infinite number of trials [105].

Furthermore, the time needed to explore this space may far exceed the typical reasoning episode duration, and reasoning episodes may explore different aspects of the space of available strategies. Thus, trials may improve phase space exploration rather than the signal-to-noise ratio [106].

### EXPERIMENTAL IMPLICATIONS

Reasoning's characteristics, particularly its lack of characteristic temporal duration, have implications at various levels. First, episodes cannot be compared in an event-related fashion. Second, defining reliable neural correlates of reasoning requires defining its characteristic temporal scales. Third, measures of brain activity should be invariant with respect to overall duration. Scaling exponents, data collapse and universality of fluctuations statistics [107-109], or explicit evolution equations for the particle's momenta and for the cross-scale fluctuation probabilities [62] can be retrieved from data and applied to unevenly lengthen trials. Thermodynamic quantities such as free energy or temperature can also be estimated for stochastic trajectories over finite time durations [97,110-113]. In all cases, the reconstruction of the underlying dynamics improves with the recording device's resolution.

Reasoning presents a dilemma between ensuring complete phase space exploration, which may require extremely long trials, and signal stationarity, which is guaranteed only for time scales much shorter than the reasoning episodes' duration. At fast time scales, the window in which relevant quantities are calculated should not introduce spurious time scales, filtering out genuine ones. Altogether, reasoning's inherently unstable nature suggests that describing it may boil down to characterizing non-stationarities and their aetiologies.

Reasoning tasks may be so difficult that only few participants manage to produce solutions within a reasonable time. This represents a shortcoming when trials are considered as independent and identically distributed, as the signal-to-noise ratio improves with the square root of the number of trials. Smoothing response times is a frequent strategy to obviate this problem, but limits or distorts the reasoning process. Furthermore, however many, short trials may insufficiently explore the phase space. Designs with few long trials may express richer spatiotemporal brain dynamics than many short ones of equivalent overall length.

Finally, while observed scaling properties may help understanding whether insight is *predictable*, i.e. whether it is an outlier or it is generated by the same distribution producing anonymous events, predicting insight onset in real data appears a challenging task, as reasoning episodes are various orders of magnitude shorter than earthquake, financial or epilepsy time series [114].

## CONCLUSIONS

Reasoning elicits an exceptionally rich repertoire of otherwise unexpressed neural properties. Its neural correlates are therefore as much helpful to neuroscientists, whom it compels to consider hitherto neglected brain properties, as they are to psychologists striving to understand its underlying processes.

Defining general and robust mechanistic properties of healthy and dysfunctional reasoning will require as yet non-standard brain metrics, experimental designs, and analytical tools (borrowed from fields rarely made available to psychologists). This may shed light on fundamental mechanistic properties of both healthy and dysfunctional reasoning, and ultimately help understanding the actions of cognitive and pharmacological interventions used as brain enhancers and targeting them to achieve desired states [115].

## ACKNOWLEDGEMENT

The author acknowledges the support of MINECO (FIS201238949-C03-01).